\documentclass[aps,pre,reprint,groupedaddress]{revtex4-1}
\usepackage{graphicx}
\usepackage{amsmath,amssymb,amsfonts}

\newcommand{\be}{\begin{equation}}
\newcommand{\ee}{\end{equation}}
\newcommand{\bea}{\begin{eqnarray}}
\newcommand{\eea}{\end{eqnarray}}

\def\d{\delta}

\def\a{\alpha}

\def\l{\lambda}

\def\s{\sigma}
\def\eps{\epsilon}
\def\bn{\bar{n}}
\def\bm{\bar{m}}
\def\bmu{\bar{\mu}}

\def\sst{\scriptscriptstyle}

\begin{document}


\title{Maximum-entropy Monte Carlo method for the inversion of the structure factor in simple classical systems}
\thanks{http://link.aps.org/doi/10.1103/PhysRevE.84.041130}

\author{Marco D'Alessandro}
\email[e-mail address: ]{marco.dalessandro@isc.cnr.it}

\affiliation{Institute for Complex Systems, National Research Council (CNR), Via del
Fosso del Cavaliere 100, 00133 Rome, Italy}


\begin{abstract}
We present a method for the evaluation of the interaction potential of an equilibrium classical system starting from
the (partial) knowledge of its structure factor. The procedure is divided into two phases both of which are based on the
maximum entropy principle of information theory.
First we determine the maximum entropy estimate of the radial distribution function constrained by the information contained
in the structure factor. Next we invert the pair function and extract the interaction potential.
The method is tested on a Lennard-Jones fluid at high density and the reliability of its results with respect to the missing information
in the structure factor data are discussed. Finally, it is applied to the experimental data of liquid sodium at 100$^{\circ}$C.
\end{abstract}

\maketitle


\section{\label{Introduction}Introduction}

The radial distribution function (RDF) of an equilibrium statistical system contains useful information concerning its
physical properties. Indeed, at least for systems governed by pairwise additive interactions, its knowledge allows one to
compute the ensemble average for observable quantities such as internal energy and pressure. Furthermore, if the same hypotheses
are satisfied, the RDF is in one-to-one correspondence with the microscopic interaction potential \cite{Henderson1974,Chayes1984} and
represents the starting point for the solution of the so-called ``inverse problem'' of statistical mechanics
\cite{Schommers1983,Reatto1986,Lyubartsev1995,Soper1996,Almarza2003,Dalessandro2010}.

Despite its central role in the analysis of a statistical system the RDF is not directly accessible from the experiments and
its estimation passes through the measurement of the structure factor. This former quantity is formally related to the RDF
by an inverse Fourier transform, which for a homogeneous and isotropic system reads:
\be \label{GR-SKdef1}
g(r) = 1 + \frac{1}{2\pi^{2}\rho} \int_{\sst{0}}^{\sst{\infty}} d k \frac{\sin(k r)}{k r} \,
k^2 \left[ S(k) - 1 \right]
\ee
where $g(r)$ and $S(k)$ are the RDF and the structure factor, respectively, and $\rho$ is the density.
So the measurement of the RDF is reduced to the evaluation of the integral appearing in Eq. \eqref{GR-SKdef1}.
Unfortunately this procedure, although conceptually correct, cannot be directly applied due to some typical limitations in the
measurement of the $S(k)$. Indeed, the experimental information is obtained by  an analysis of the x-ray and/or neutron
diffraction data. These techniques provide results over a finite $k$ range and a number of nontrivial corrections on measured
data are needed. So the resulting experimental structure factor turns out to be incomplete and typically spoiled by systematic and
statistical errors. As a consequence, the RDF obtained by means of Eq.1 \eqref{GR-SKdef1} may present non physical features and
spurious structures could emerge.

In order to overcome these difficulties different approaches have been pursued. A promising class of solutions is provided by simulation
assisted methods in which a molecular dynamics or a Monte Carlo (MC) simulation is driven with the aim to minimize the differences between
the simulated structure factor and the experimental data.
Among the results belonging to this class we cite the reverse Monte Carlo technique, proposed by McGreevy and Pusztai in \cite{McGreevy1988},
which implements the transition probability on the basis of the $\chi^{2}$ function between the reference and the simulated structure factors;
this procedure, however, does allow one to determine the pair interaction potential. Further approaches are provided by the method proposed by T\'{o}th
\cite{Toth2001} and based on the previous work of Lyubartsev and Laaksonen \cite{Lyubartsev1995}, and by the solution due to Almarza, Lomba, and Molina
\cite{Almarza2004}. These methods consist in an iterative procedure for the evaluation of an effective pair potential compatible with the experimental
data, so they attempt to provide a true solution of the inverse problem starting from the structure factor. A comprehensive review of the simulation assisted
methods is given in \cite{Toth2007}.

Since we are dealing with the reconstruction of the RDF starting from the partial knowledge of the experimental $S(k)$ one question concerning the
uniqueness of the solution naturally arises; at the same time it is desirable that no information besides that contained in the structure factor
is transferred to the RDF during the reconstruction.
Both of these issues can be addressed using the maximum entropy principle (ME) \cite{Jaynes1957} as the guideline for the definition of the
reconstruction procedure. Indeed ME has the remarkable feature of producing the highest entropy solution compatible with the given constraints,
so the corresponding estimate for the RDF is ``maximally noncommittal with regard to the missing information'' \cite{Jaynes1957}.
ME-based algorithms for the inversion of the structure factor were first developed by Root, Egelstaff and Nickel \cite{Root1986} and by Soper
\cite{Soper1986}. In these papers it has been shown that the adoption of ME improves the Fourier transform of the structure factor data and reduces
the spurious structure in the RDF.
ME has been introduced for the first time in contest of the inverse problem by Cilloco in \cite{Cilloco1993}; the method described in this paper
used ME inside a Monte Carlo simulation scheme. It has been shown that a  maximum entropy ensemble of configurations compatible with a given reference
RDF can be built adopting a suitable definition of the transition probability between neighbor states. This approach has been recovered and extended
in \cite{Dalessandro2010}; the transition probability has been reinterpreted as an information-based feedback controller and an ``integral term'' has
been added. The authors evidenced that this quantity converges to the interaction potential thus providing a ME-based solution of the inverse problem.

The purpose of this paper is to present a ME Monte Carlo method for the inversion of the experimental structure factor. The procedure
is mainly based on the statistical properties of the pair distribution functions both in the $r$ and in the $k$ space. We will show that,
thanks to the above mentioned features of ME, the algorithm provides a reliable reconstruction of the RDF starting from a limited
knowledge of the experimental structure factor. Once the $S(k)$ has been inverted, we can apply the technique described in \cite{Dalessandro2010}
to the resulting RDF and extract the (pair) interaction potential.

The paper is organized as follows. Section \ref{Theory} contains a detailed theoretical description of our procedure. In Sec. \ref{Applications}
we test the method for a Lennard-Jones fluid assuming different cutting points of the input $S(k)$ and we invert the experimental data of liquid sodium
at 100$^{\circ}$C.  Finally, in Sec. \ref{Discussion} we discuss our results and present some concluding remarks.

\section{\label{Theory}Theory}

We present a statistical description of a simple monoatomic fluid, in an analogous way of what has been done in \cite{Dalessandro2010}, and extend
this analysis to the Fourier transform of the RDF. Then we describe a procedure for the construction of a ME ensemble of configurations
constrained by the (partial) knowledge of the structure factor.

\subsection{\label{Preliminaries}Preliminaries}

We define the notion of a \emph{model} system, which is a homogenous and isotropic collection of pointlike elements with average density $\rho$.
Given an arbitrary configuration $\bf{x}$ of the model we define two quantities that will be relevant for the subsequent analysis: the local
sampling of the elements pair function (PF) $n$ and its Fourier transform $\bn$. The former quantity provides the number of particles $n_{i}$
inside the $i$th spherical shell of width $\d r$ centered on a reference element; the sampling is performed up to the maximum value $r_{M}$ and
consequently the index $i$ runs from 1 to $N=r_{M}/\d r$. The Fourier transform (FT) of the local PF is defined through the equation
\be \label{FTdef1}
\bn_{j} = {\cal F}_{j}\left(n\right) = \sum_{i=1}^{N} \frac{\sin(k_{j} r_{i})}{k_{j} r_{i}} \, n_{i}
\, , \quad
j = 1,..,N
\ee
where $r_{i}$ is the value of the radius associated to the $i$th shell: $r_{i} = i\d r$. Given the local pair function $n$, its FT $\bn$ represents
an $N$ elements vectors in the $k$ space. The component $\bn_{j}$ contains the $k$-space value at $k_{j} = j \d k$; the sampling is performed with a uniform
step of width $\d k$, chosen according to the relation
\be \label{dkdef1}
\d r \d k = \frac{\pi}{N}
\ee
We also introduce the notion of inverse Fourier transform (IFT). Given a $k$-space vector $\bn$ we define its IFT through the equation
\be \label{IFTdef1}
n_{i} = {\cal F}^{\sst{(-1)}}_{i}\left(\bn\right) =
\frac{2}{N} \sum_{j=1}^{N} r_{i}^{2}k_{j}^{2}\,\frac{\sin(k_{j} r_{i})}{k_{j} r_{i}} \, \bn_{j}
\ee
Equation \eqref{dkdef1} ensures the orthonormality of the discrete basis of functions adopted in Eqs. \eqref{FTdef1} and \eqref{IFTdef1}, namely,
\be \label{FTorthogonality1}
\sum_{j=1}^{N} \sin(k_{j} r_{i}) \sin(k_{j} r_{l}) = \frac{N}{2} \d_{il}
\ee
so the transformation of a PF from the $r$ space to the $k$ space and back again will reproduce the initial function \cite{Lado1971}.
It is worth mentioning that, according to Eq. \eqref{dkdef1}, a sampling of width $\d r$ in the $r$ space produces a $k$ space
vector with a maximum wave number given by $k_{M} = \pi / \d r$, in agreement with the Shannon-Nyquist sampling theorem \cite{Shannon1949}.

Since we are interested in the construction of the average pair functions (both in the $r$ and $k$ space) we have to extend the notion of
local PF and of its FT to a large number of configurations. So we introduce a probability function $p(\bf{x})$ defined upon the model
configuration space and we collect an ensemble of $s$ elements extracted according to $p$.
The global samplings over the ensemble are defined as the average values of the local ones:
\bea \label{mbmdef1}
m_{i} & = & \frac{1}{s} \, \sum_{\a = 1}^{s} n_{i}^{(\a)} \nonumber \\
\bm_{j} & = & {\cal F}_{j}\left(m\right) = \frac{1}{s} \, \sum_{\a = 1}^{s} \bn_{j}^{(\a)}
\eea
where the index $\a$ labels the elements of the ensemble and the last equality holds due to the linearity of the FT.
The radial distribution function and the structure factor of the model system are defined starting from the global quantities
\eqref{mbmdef1} in the limit $s \rightarrow \infty$. The RDF is obtained by normalizing the global PF built on the model ensemble
with the pair function of a uniform reference system (perfect gas) with the same density of the model one:
\be \label{RDFdef1}
g(r_{i}) = \frac{m_{i}}{m_{i}^{\sst{(pg)}}}
\ee
where $m_{i}^{\sst{(pg)}} = 4\pi\rho r_{i}^{2} \d r$ is the perfect gas pair function.
The structure factor is defined in terms of the FT of the global pair function via the relation
\be \label{SKdef1}
S(k_{j}) = 1 + \bm_{j} - \bm^{\sst{(pg)}}_{j}
\ee
This definition provides the usual notion of $S(k)$ for an isotropic statistical system, indeed making use of Eqs.
\eqref{FTdef1} and \eqref{RDFdef1} and performing the continuum limit gives
\be \label{SKdef2}
S(k) = 1 + 4\pi\rho \int_{\sst{0}}^{\sst{r_{M}}} d r \frac{\sin(k r)}{k r} \, r^2 \left[ g(r) - 1 \right]
\ee
which is the formal definition of structure factor adopted in statistical mechanics. We observe that, due to the finite size of the model
system, the integral in Eq. \eqref{SKdef2} extends up to the maximum sampled value of the model RDF. Consequently, according to the
Shannon-Nyquist sampling theorem, the maximum allowed $k$ resolution is given by $\pi / r_{\sst{M}}$.

We conclude this preliminary section by introducing a useful notation for dealing with the Fourier transforms.
Since the FT and its inverse are realized as linear combinations among the $\bn$ and the $n$ variables, respectively,
we can write
\be \label{FTdef2}
\bn_{j} = \sum_{i=1}^{N-1} c_{ij} \, n_{i} \, , \qquad
n_{i} = \sum_{j=1}^{N-1} c^{\sst -1}_{ij} \, \bn_{j}
\ee
where according to Eqs. \eqref{FTdef1}, \eqref{dkdef1} and \eqref{IFTdef1} both $c_{ij}$ and its inverse are symmetric
matrices given by
\be \label{c&cm1def1}
c_{ij} = \frac{N}{\pi}\frac{\sin\left[\left(\frac{\pi}{N}\right)ij\right]}{ij} \, \quad
c^{\sst -1}_{ij} = 2\pi\frac{\sin\left[\left(\frac{\pi}{N}\right)ij\right] ij}{N^{2}}
\ee
The sum in Eqs. \eqref{FTdef2} has been restricted to the first $N-1$ elements since the last one gives a zero contribution
for algebraic reason related to the definition of the matrices \eqref{c&cm1def1}. We point out that Eq.
\eqref{FTorthogonality1} ensures that the matrix product of these quantities gives the identity matrix as expected.

\subsection{\label{model distribution}Analysis of the model distribution function}

We are interested in computing the probability distribution of the FT of the global pair function built on the model ensemble.
So we suppose that an ensemble of configurations has been extracted according to a given probability distribution
$p(\bf{x})$ and we compute the probability associated to a particular sampling $\bm$ as a function of the parameters of the
underlying ensemble distribution.

To achieve this task we start from the probability of the local sampling of the pair function $n$. The $i$th shell of the PF
follows a Poisson distribution with expectation value $\mu_{i}$ \cite{Dalessandro2010}; we assume that the system exhibits a hard
core (HC) structure with radius $r_{\sst{0}}$ so that the expected number of particles $\mu_{i}$ is zero for $i$ lower than the
threshold value $N_{\sst{0}} = r_{\sst{0}} / \d r$ and is strictly positive otherwise.
Since there is no correlation among different shells the complete distribution is obtained as the product of the single shell
values and reads
\be \label{poissondef1}
{\cal P}_{\mu}(n) = \prod_{i=N_{\sst{0}}}^{N} e^{-\mu_{i}}
\frac{(\mu_{i})^{n_{i}}}{n_{i}!}
\ee
The HC structure of the reference distribution imposes that ${\cal P}_{\mu}$ is zero if there is some $n_{i}>0$ for $i<N_{\sst{0}}$.
The FT of the local sampling of the pair function $\bn$ is defined through a linear combination of the $n$ variables \eqref{FTdef2},
so its expectation value and its covariance can be expressed in terms of the $\mu_{i}$ parameters:
\bea \label{bnExpValue1}
\textrm{E}\left(\bn_{j}\right) &=& \bmu_{j} = \sum_{i} c_{ij}\mu_{i} \nonumber \\
\textrm{Cov}\left(\bn_{j},\bn_{k}\right) &=& \xi_{jk} = \sum_{i} c_{ij}c_{ik}\mu_{i}
\eea
We observe that, due to the linear combination \eqref{FTdef2}, the covariance matrix of the $\bn$ variables is not diagonal even if the
original variables $n$ are uncorrelated. The variable $\bm$ is defined as the average of the $\bn^{(\a)}$ \eqref{mbmdef1}, so it has
the same expectation value $\bmu$ and a covariance given by $\xi / s$. Its asymptotic distribution can be computed using a multivariate
central limit theorem; this theorem states that the distribution function of the reduced variable $\sqrt{s}(\bm_{j}-\bmu_{j})$
converges, in the limit $s \rightarrow \infty$, to a multivariate Gaussian with zero mean and covariance given by $\xi$. So we have
\be \label{bmProbability2}
\sqrt{s}(\bm-\bmu)
\underset{{\sst s \gg 1}}{\sim}
{\cal N}_{{\sst 0}}
\ee
where:
\be \label{bmProbability3}
{\cal N}_{{\sst 0}}\left(\bf{x}\right) =
\frac{1}{(2\pi)^{\sst N/2} |\xi|^{\sst 1/2}} \,
e^{-\frac{1}{2} \bf{x}^{T} (\xi)^{\sst -1} \bf{x}}
\ee
is the multivariate distribution function with zero mean and $|\xi|$ represents the determinant of the covariance matrix. It turns out that
the elements of $\bm$ are linear dependent and consequently the covariance matrix is singular. This is due to the fact that only  the nonzero
components of the local PF contribute to the linear combination \eqref{FTdef2}, so the number of independent elements of $\bm$ is $N-N_{\sst{0}}$.
In order to avoid  a singular covariance matrix we have to restrict our analysis to a set of independent elements of $\bm$; in this domain the
covariance matrix can be inverted and its inverse reads:
\be \label{InverseCOV1}
\xi^{\sst -1}_{jk} = \sum_{i=N_{\sst{0}}}^{N-1}
\tilde{c}^{\sst -1}_{ij}\tilde{c}^{\sst -1}_{ik}\frac{1}{\mu_{i}}
\ee
where the indices $j,k$ run from 1 to $N-N_{\sst{0}}$ and the tilde indicates that the matrices are restricted to the subset of
independent variables.

This analysis shows that the asymptotic distribution for the independent subset of the $\bm$ components
is described by a multivariate Gaussian distribution ${\cal N}_{\bmu}$ with mean $\bmu$ and inverse covariance $s \cdot \xi^{\sst{-1}}$.
We observe that both the expectation value \eqref{bnExpValue1} and the inverse covariance matrix \eqref{InverseCOV1} are functions of the
parameters of the original distribution function \eqref{poissondef1}.

\subsection{\label{Max Ent approach}Maximum entropy approach to the inverse problem}

We consider a monoatomic system and assume that for a given density $\rho$ and temperature $T$ the structure factor $S_{t}(k)$ of the
system is known up to the maximum value $k_{\sst{M}}$. We will refer to this system as the \emph{target}.

The aim of this section is to define a procedure for the evaluation of an \emph{equilibrium} model distribution function $p(\bf{x})$
compatible with the information contained in the target structure factor.
The method is based on the maximum entropy principle and is realized inside a Monte Carlo simulation scheme.
MC represents an effective tool to pursue this approach: the maximization of configurational entropy is produced by the MC random movements
for the construction of the trial configurations (source of entropy) while the transition probability among neighbor states selects the configurations
and acts as a source of information. At equilibrium these two mechanisms are in balance, the net amount of information loss is zero, and the system
approaches a state of maximum entropy consistently with the given constraints.

The main advantage of this kind of procedure is that the ME solution is sought in terms of a ``real'' physical system which possesses a true configuration
space beyond the two-body pair function; so its equilibrium distribution implicitly defines the correlation functions of any order.
Inside this scheme, the implementation of the ME algorithm realizes the maximization of the whole configurational entropy and not only of the
two-body contribution. This method provides the maximum entropy estimate of the complete equilibrium distribution of the model system and the ensemble of
configurations built according to it can be used to compute the average value of any quantity of interest.

Since, under this perspective, the transition probability is the natural object in which the knowledge on the system is codified, we seek this quantity
with the aim of building a model distribution function that produces an expectation value of $\bm$ consistent with the target reference value $\bmu_{t}$
(the proper definition of this parameter on the basis of the target $S(k)$ will be discussed in the next section).
To achieve this task we make use of the method developed in \cite{Cilloco1993,Dalessandro2010} and we maximize the log-likelihood function between the
model pair function and the target reference value. This choice is based on statistical reasons: in the limit of a large number of configurations the
average $\bm$ computed over the model ensemble converges to the expectation value $\bmu$ of the model distribution function and the log-likelihood can
be related to the relative entropy $D$ (Kullback-Leibler divergence \cite{Kullback1951}) between the model and the target distributions:
\be \label{LikelihoodRelativeEntropy1}
\ln {\cal N}_{\bmu_{t}}\left(\bm\right) = -
D \left( {\cal N}_{\bmu} || {\cal N}_{\bmu_{t}} \right)
\ee
so the maximization of the log-likelihood function is asymptotically equivalent to the minimization of the relative entropy \eqref{LikelihoodRelativeEntropy1}.
Given two distributions  $p$ and $q$, the relative entropy $D(p||q)$ is positive definite and vanishing only if $p=q$, so a complete maximization of the
likelihood function implies the equality of the distributions.

So our general strategy is the following: we perform a MC simulation using a transition probability which maximizes the log-likelihood function defined above.
This procedure builds a maximum entropy ensemble of configurations constrained by the target $S(k)$ and the radial distribution function computed over this ensemble
is the maximum entropy estimate of the inverse Fourier transform of the target structure factor.
Since the maximum entropy principle has the feature of providing reliable estimates on the basis of a partial input of information, we expect that this procedure
should be able to produce a correct reconstruction of the radial distribution function starting from a limited knowledge of the structure factor.

In the next section we will describe some details of the implementation of this procedure. The applications of the method and
some checks of its reliability and sensitivity to the amount of missing information are discussed in Sec. \ref{Applications}.

\subsection{\label{max log-likelihood}Maximization of the log-likelihood function}

Assume that the $S_{t}(k)$ has been measured with a resolution $\d k$ up to the value $k_{M}$; so the target input is given by $N_{t} = k_{M} / \d k$
samplings of the structure factor.

The first step consists in a proper definition of the model system (see Sec. \ref{Preliminaries}): the value of the model density is chosen equal
to the target one and the model pair function is sampled up to the maximum value $r_{M} = \pi / \d k$. This choice ensures that the model structure
factor is sampled with the same resolution as the one of the target system.
The spatial resolution $\d r$ in the model configuration space is chosen with the double task of producing an accurate sampling of the model RDF and
to ensure that the maximum sampling value of the model structure factor (given by $\pi / \d r$) is greater than the target value $k_{M}$.

We define the procedure for the construction of the model ensemble based on the maximization of the log-likelihood function described in the previous section.
First we build the reference distribution on the basis of available information concerning the target structure factor [see Eq. \eqref{SKdef1}]:
\be \label{bmutdef1}
\bmu_{t\, j} = \bm^{\sst{(pg)}}_{j} + S_{t}(k_{j}) - 1
\ee
where $j$ extends over all the shells in the model system (from 1 to $N=r_{M} / \d r$) and we impose that $S_{t}(k_{j})$ is equal
to 1 for all $j > N_{t}$.
Given the $\bmu_{t}$ we compute its inverse Fourier transform $\mu_{t}$ which represents the expectation value of the target pair function.
Due to the lacking information in the target structure factor the $\mu_{t}$ provides a \emph{biased} reconstruction of the true target pair function; typically
this function exhibits nonphysical behavior such as, for instance, strong oscillations inside the hard core radius.

Next we analyze the construction of the log-likelihood ratio. Assume that we have performed $s$ MC iterations. For each point of the path we compute a local
sampling of the PF $n^{(\a)}$ and its FT $\bn^{(\a)}$ and we construct the global pair functions $m$ and $\bm$.
Then we select a reference particle and compute a local sampling of the PF $n^{\sst{(1)}}$; at the same time the particle is randomly moved and the new local
sampling of the PF is stored in the array $n^{\sst{(2)}}$. In this way we obtain two different samplings of $\bm$ at the level $s+1$, namely
$\bm^{\sst{(1)}}$ and $\bm^{\sst{(2)}}$. Then we perform a \emph{cut} in the model system consistently with the missing information in the target reference
function: so we substitute the perfect gas value in both the $\bm$ samplings for $j > N_{t}$. This procedure provides $\bm^{cut \, \sst{(1)}}$ and
$\bm^{cut \, \sst{(2)}}$ which are the natural quantities to be compared with $\bmu_{t}$.
Finally we define the log-likelihood ratio via the relation
\be \label{Likelihoodratio1}
\d \l = \ln \frac{{\cal N}_{\bmu_{t}}\left(\bm^{cut \, \sst{(1)}}\right)}
{{\cal N}_{\bmu_{t}}\left(\bm^{cut \, \sst{(2)}}\right)}
\ee
The transition probability selects all the trial configurations which maximize the likelihood function $(\d \l < 0)$ \footnote{In this paper we follow
the method introduced in \cite{Cilloco1993} for the maximization of the likelihood function.} and consequently the distribution of the $\bm^{cut}$ converges
to a multivariate Gaussian defined by the target parameters $\bmu_{t}$ and $\mu_{t}$ [see Eqs. \eqref{bnExpValue1} and \eqref{InverseCOV1}].
At the same time the model global sampling $m$ converges to the \emph{unbiased} reconstruction of the target RDF and its FT $\bm$ builds the complete target
structure factor. So, thanks to the ME approach, we build a complete estimate of the target distribution function on the basis of a limited amount of information.

We conclude this section by analyzing the expression of the log-likelihood ratio. In the limit of a large number of configurations we can expand Eq.
\eqref{Likelihoodratio1} in power of $s$. The leading order contribution reads:
\be \label{Likelihoodratio2}
\d \l = \sum_{i,j=1}^{N_{t}} \left(\bn_{i}^{cut \, \sst{(2)}} - \bn_{i}^{cut \, \sst{(1)}} \right)
\xi^{\sst -1}_{ij}\left(\bm_{j}^{cut}-\bmu_{t\,j}\right)
\ee
Equation \eqref{Likelihoodratio2} evaluates the log-likelihood ratio as the weighted sum of the differences between the actual and the trial local sampling $\bn$.
The weights are proportional to the discrepancy between the global $\bm^{cut}$ and the target reference function and, due to the non diagonal correlation matrix,
each term in the sum depends on the whole difference $(\bm^{cut}-\bmu_{t})$.
It is interesting to recast this equation in terms of the local PFs in the $r$ space; to obtain this result we make use of Eqs. \eqref{FTdef2} and
\eqref{InverseCOV1}; this provides:
\be \label{Likelihoodratio3}
\d \l = \sum_{i=N_{\sst{0}}}^{N} \left(n_{i}^{\sst{(2)}} - n_{i}^{\sst{(1)}} \right)
\frac{m_{i}^{\sst{(bias)}} - \mu_{t\, i}}{\mu_{t\, i}}
\ee
where $m^{\sst{(bias)}}$ represents the inverse Fourier transform of $\bm^{cut}$.
We see that, once reformulated in the configuration space, the log-likelihood becomes \emph{diagonal}: the $i$th term in the sum \eqref{Likelihoodratio3}
is weighted by the $i$th shell value of the discrepancy between the model and the target global (biased) PF. This behavior is a consequence of the
independence between the local pair functions related to different shells [see Eq. \eqref{poissondef1}].
It is worth mentioning that Eq. \eqref{Likelihoodratio3} is strongly reminiscent of the log-likelihood ratio computed in \cite{Cilloco1993,Dalessandro2010}
starting from the knowledge of the target RDF. Indeed we recognize that the weighted difference between $m^{\sst{(bias)}}$ and $\mu_{t}$ is the first order
expansion \footnote{The expansion of the logarithm is due to the use of a central limit theorem, performed in section \ref{model distribution}, for the determination
of the probability distribution of $\bm$.} of $\ln (m^{\sst{(bias)}}/ \mu_{t})$.
Furthermore, if the complete target structure factor is provided, then the input of information content becomes equivalent to the knowledge of the target RDF;
in this case $\mu_{t}$ represents the true value of the target PF and the $m^{\sst{(bias)}}$ coincides with the model global PF $m$ thus providing an identical
expression of the log-likelihood ratio.

\subsection{A remark on the transition probability}

In Sec. \ref{max log-likelihood} we stated that the transition probability is defined in a way to accept all the trial configurations with a
log-likelihood ratio lower than zero $(\d \l < 0)$. In order to better comprehend the reasons behind this choice
it is useful to briefly recall the main results concerning the analysis of the transition probability described in \cite{Dalessandro2010}.
Following the approach outlined in this reference we can interpret Eq. \eqref{Likelihoodratio3} has a \emph{proportional feedback controller}
which selects the model system configurations on the basis of the ``error'' $e = (m^{\sst{(bias)}} - \mu_{t})/\mu_{t}$ between the target and the
model global pair function.
This interpretation suggests the formulation of an improved expression for $\d \l$, based on the theory feedback systems, that also include
an \emph{integral term} apart from the proportional one; this latter quantity keeps into account all the errors in the steps preceding the
actual one. In this way we realize a proportional-integral controller, schematically defined as:
\bea
\d \l = \sum_{i} \left(n_{i}^{\sst{(2)}} - n_{i}^{\sst{(1)}} \right)
\left( k_{p} e_{i} + k_{I} \sum_{\a} e_{i}^{(\a)} \right) \nonumber
\eea
where $k_{p}$ and $k_{I}$ are the coefficients of the proportional and integral term, respectively. The transition probability is defined as
$\min\{1,\exp(-\d \l)\}$ and the proportional and integral coefficients are fixed with the aim to ensure the correct fluctuation of the model PF
around its average value. Results reported in \cite{Dalessandro2010} evidence that this approach allows one to build the correct equilibrium distribution
of the target system; furthermore, the interaction potential emerges as the asymptotic limit of the integral term.

The procedure delineated above can be applied in the present case and would allow one to directly extract the interaction potential from the knowledge of the
structure factor. Instead, we have adopted a different implementation of the feedback controller in which the integral term is absent and the proportional
coefficient is virtually infinite: so only the trial configurations with a log-likelihood ratio lower than zero are accepted.
The main advantage of this choice is that a pure proportional controller ensures a straightforward and stable convergence of the inversion procedure, so the RDF
is obtained without the need of setting any parameters. In this way we can perform an intermediate check of the inversion procedure.
Obviously, the extraction of the interaction potential requires the subsequent inversion of the RDF using the method described in \cite{Dalessandro2010}.

\section{\label{Applications}Applications}

The technique previously described has been applied to a simple Lennard-Jones fluid and to the experimental structure factor data of the liquid
sodium at 100$^{\circ}$C measured by Greenfield, Wellendorf and Wiser in \cite{Greenfield1971}.

The inverse MC simulation is realized in the $NVT$ ensemble: the model configuration space is a cubic volume of linear length $L$  with $N_{p}$
pointlike particles and the periodic boundary conditions together with the minimum image convention have been adopted.

The transition probability between neighbor states has been evaluated by using Eq. \eqref{Likelihoodratio3} for the computation of the
log-likelihood ratio. This formula requires knowledge of the HC radius which is \emph{a priori} unknown; a brief estimate of its value can be
obtained (as suggested by Reatto in \cite{Reatto1986}) by computing the inverse FT of the structure factor and by taking a fraction of the
$r$ position of its first peak.
ME will allow this estimate to be corrected to its optimal value during the simulation.

\subsection{\label{ResultLJ}Results for the Lennard-Jones system}

\begin{figure*}
  \begin{center}
    \begin{tabular}{cc}
      \resizebox{80mm}{!}{\includegraphics{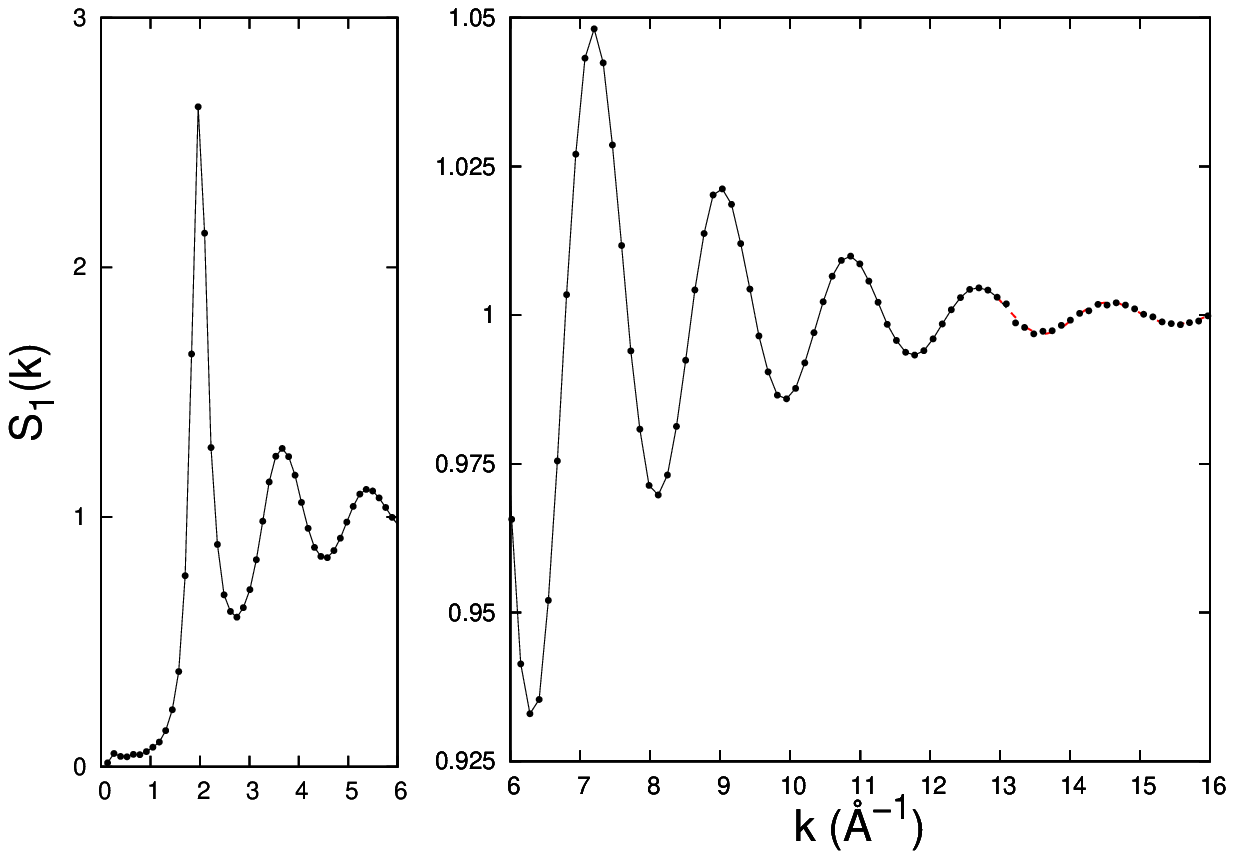}} &
      \resizebox{80mm}{!}{\includegraphics{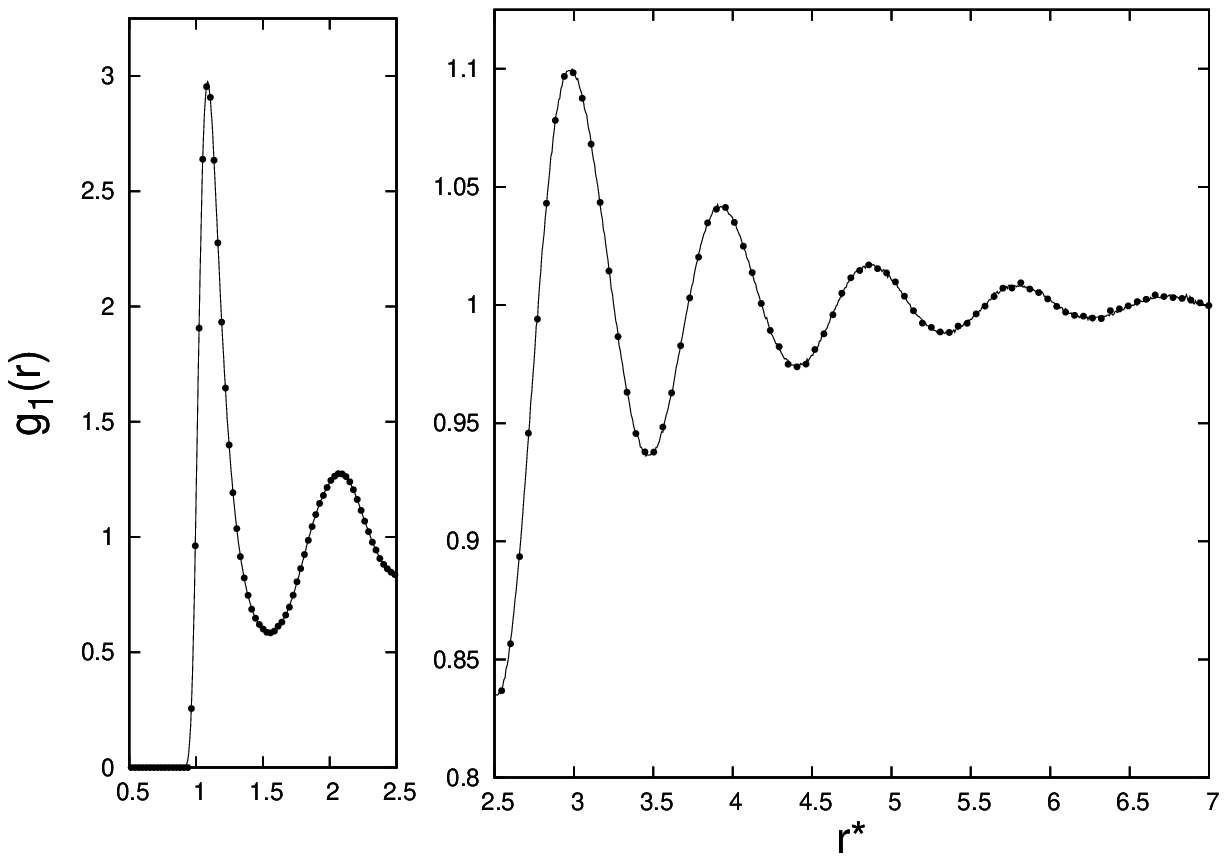}} \\
      \hspace{9.5cm} (a) \\
      \resizebox{80mm}{!}{\includegraphics{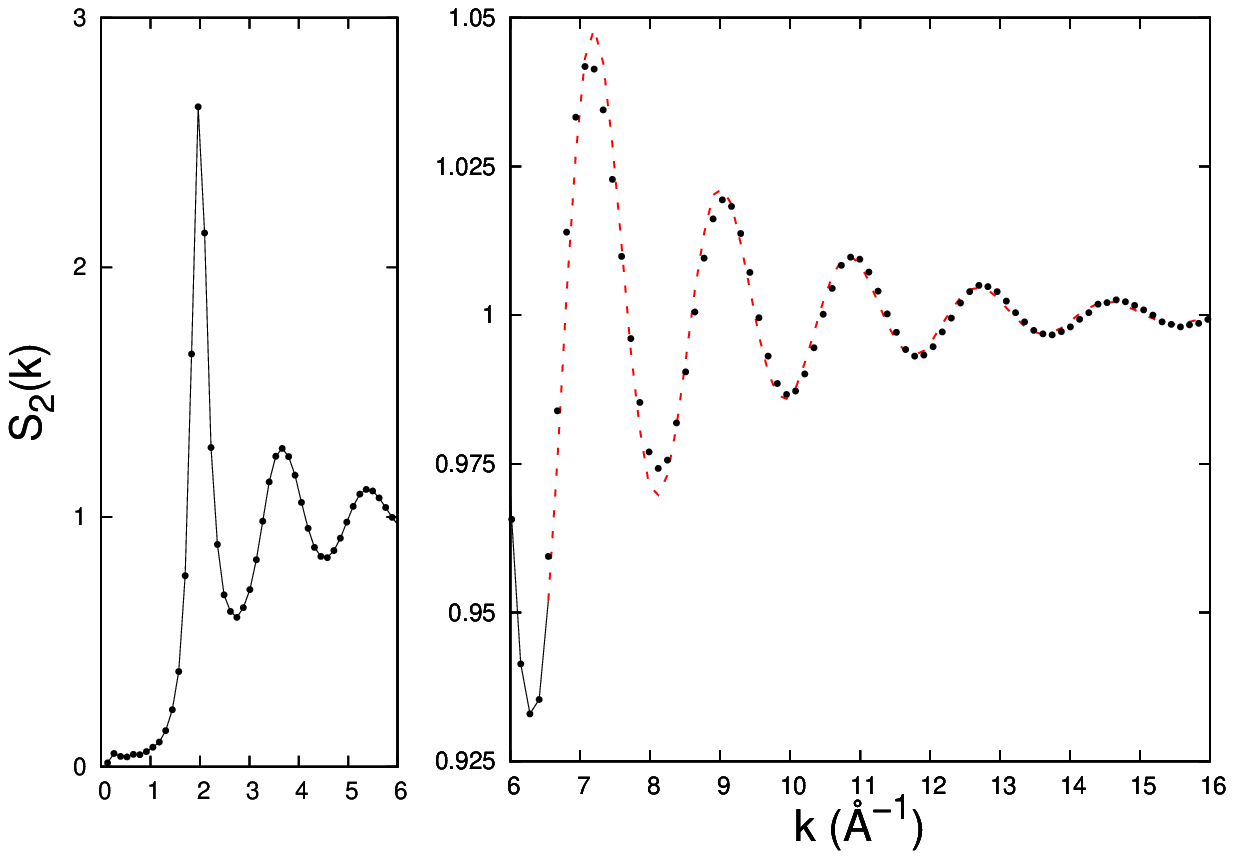}} &
      \resizebox{80mm}{!}{\includegraphics{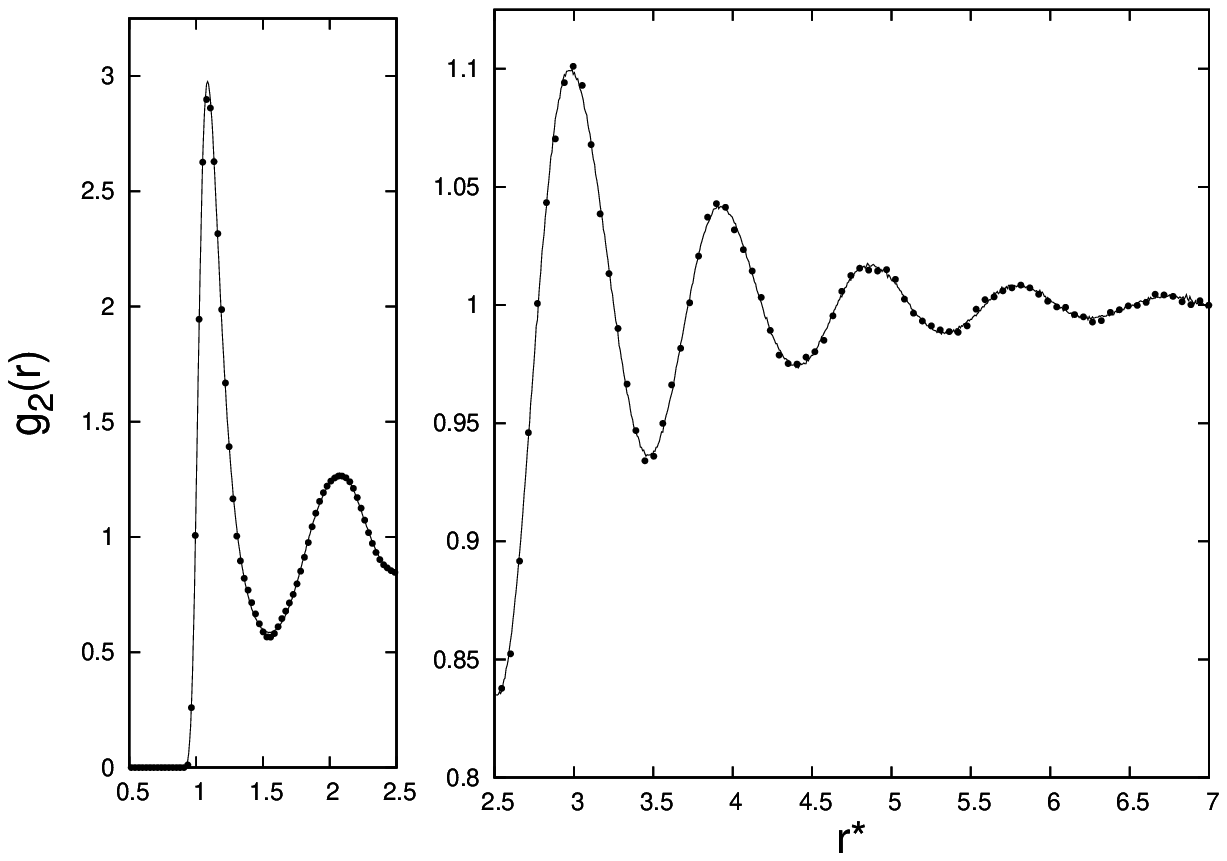}} \\
      \hspace{9.5cm} (b) \\
      \resizebox{80mm}{!}{\includegraphics{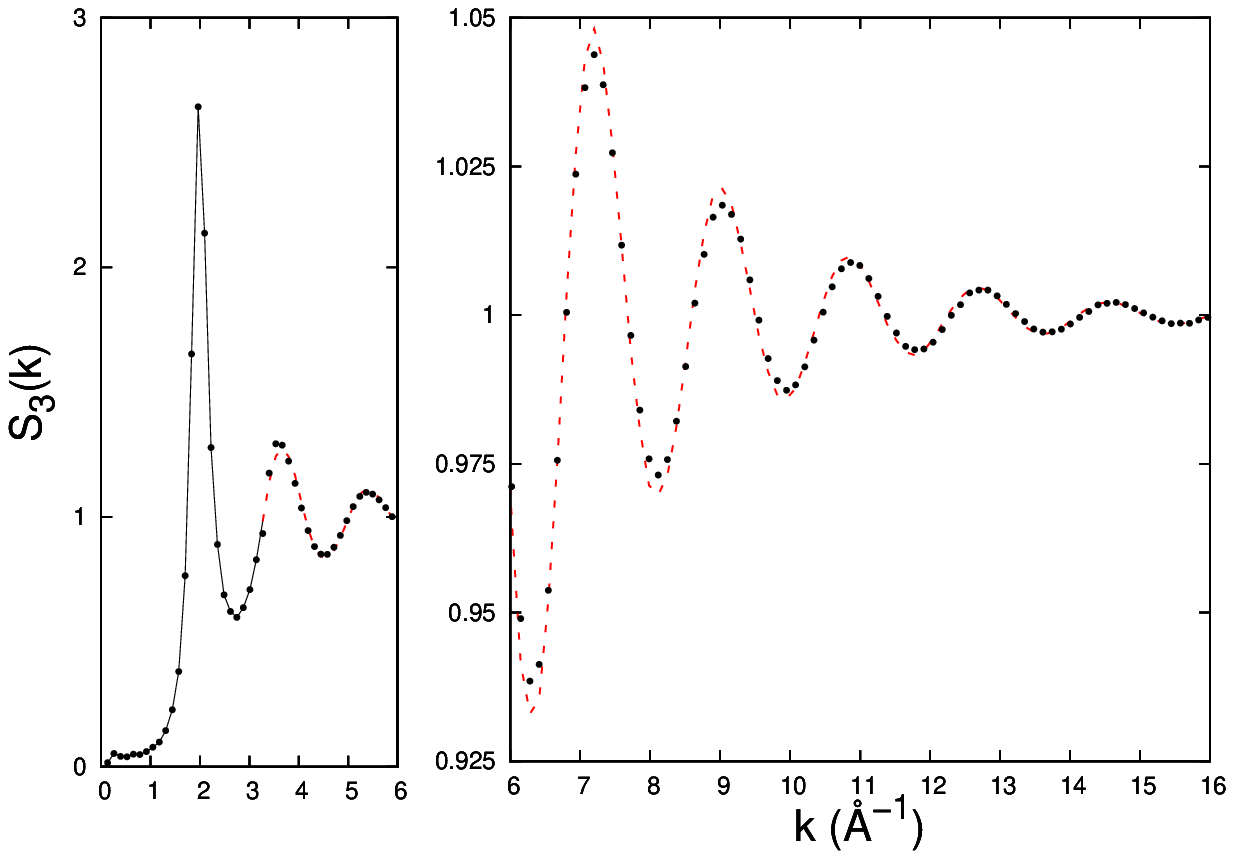}} &
      \resizebox{80mm}{!}{\includegraphics{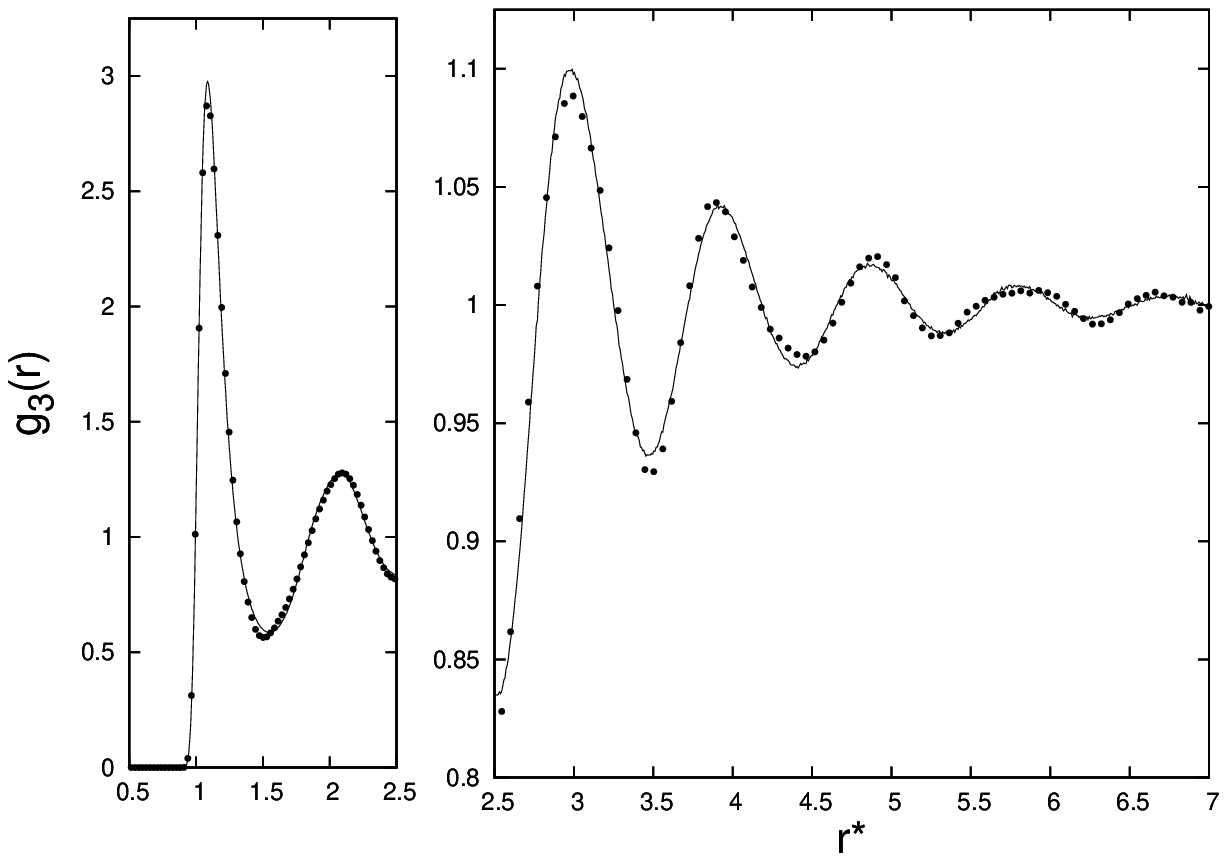}} \\
      \hspace{9.5cm} (c) \\
    \end{tabular}
    \caption{(Color online) Results of the inversion procedure for a Lennard-Jones system. The left column contains the plots of the structure factor:
    continuous line for the target $S(k)$ used as model input, dotted line for the target $S(k)$ in the $k$ region beyond $k_{M}$, and filled circles
    for the model $S(k)$. The right column contains the radial distribution functions: continuous line for the target RDF and filled circles
    for the model RDF. (a) Target $S(k)$ truncated at 13 \AA ${}^{-1}$; (b) target $S(k)$ truncated at 6.5 \AA ${}^{-1}$; and (c) target $S(k)$ truncated
    at 3.2 \AA ${}^{-1}$.}
    \label{SK-GR LJ}
  \end{center}
\end{figure*}

We consider a system described by the Lennard-Jones potential with argon-like parameters $\s=3.405$ \AA\ and $\eps/k_{B}=119.76$ K.
The target structure factor is evaluated by performing a Metropolis MC simulation on a system of 864 particle at the reduced density
$\rho^{*}=\rho \s^{3}=0.84$ and reduced temperature $T^{*}=k_{B}T/\epsilon=0.75$, near the triple point.
The simulation run for $2\times 10^4$ cycles after equilibration. The $g(r)$ has been evaluated up to $r^{*}=r/\s=7.05$ (24 \AA), the width
of the shells for the measure of the $g(r_{i})$ was $\d r = 2.4\times 10^{-2}$ \AA, and the number of measured points was $10^3$. The structure
factor has been evaluated using the procedure described in Sec. \ref{Preliminaries}; the $k$ resolution is given by Eq. \eqref{dkdef1}
and is equal to $\d k = 0.13$ \AA ${}^{-1}$.

Once the target $S(k)$ was computed we performed the inversione procedure for the reconstruction of the RDF. In order to check the sensitivity
of this approach we truncated the target $S(k)$ at different values of $k$ and we proceeded to the reconstruction for each of the truncated function.
So we built three structure factors, namely, $S_{t\,\sst{1}}(k)$ (truncated at $k_{M} = 13$ \AA ${}^{-1}$), $S_{t\,\sst{2}}(k)$ (truncated at $k_{M} = 6.5$
\AA ${}^{-1}$), and $S_{t\,\sst{3}}(k)$ (truncated at $k_{M} = 3.2$ \AA ${}^{-1}$).
Then the reconstruction procedure started for $2\times 10^4$ cycles after equilibration. In this way we produced three radial distribution functions,
namely, $g_{\sst{1}}(r)$, $g_{\sst{2}}(r)$, $g_{\sst{3}}(r)$ and the corresponding structure factors  $S_{\sst{1}}(k)$, $S_{\sst{2}}(k)$ and $S_{\sst{3}}(k)$.

Results are reported in Fig. \ref{SK-GR LJ}.
The first line contains the outcomes of the inversion starting from $S_{t\, \sst{1}}(k)$. The maximum difference between the target and the
model structure factor for $k$ up to $k_{M}$ is about $4\times 10^{-4}$ and the procedure reconstructed the target $S(k)$ for $k > k_{M}$ with
an error lower than $1\times 10^{-3}$; the model RDF reproduces the target values with a maximum difference of about $2\times 10^{-2}$.
The second line reports results obtained using the information content of $S_{t\, \sst{2}}(k)$. Even in this case the maximum discrepancy up
to $k_{M}$ (6.5 \AA ${}^{-1}$) is about $4\times 10^{-4}$ and the procedure reconstructed the target structure for $k > k_{M}$
with an error lower than $1\times 10^{-2}$; the maximum difference between the target and the model RDF is about $5\times 10^{-2}$.
Finally, in the last line of Fig. \ref{SK-GR LJ} we present the results of the inversion of $S_{t\, \sst{3}}(k)$. The maximum difference between
the target and the model structure factor up to $k_{M}$ (3.2 \AA ${}^{-1}$) is about $4\times 10^{-4}$ as in the previous cases and, despite
the modest information content of the target structure factor, the model $S(k)$ reproduces the target one for $k > k_{M}$ with an error lower than
$4\times 10^{-2}$. The corresponding RDF reconstructs the target function with a maximum discrepancy of about $6\times 10^{-2}$.
This analysis evidences the effectiveness of the maximum entropy principle to provide accurate reconstructions on the basis of a limited amount of
information.

\begin{figure}
\includegraphics[width=9cm,height=8cm]{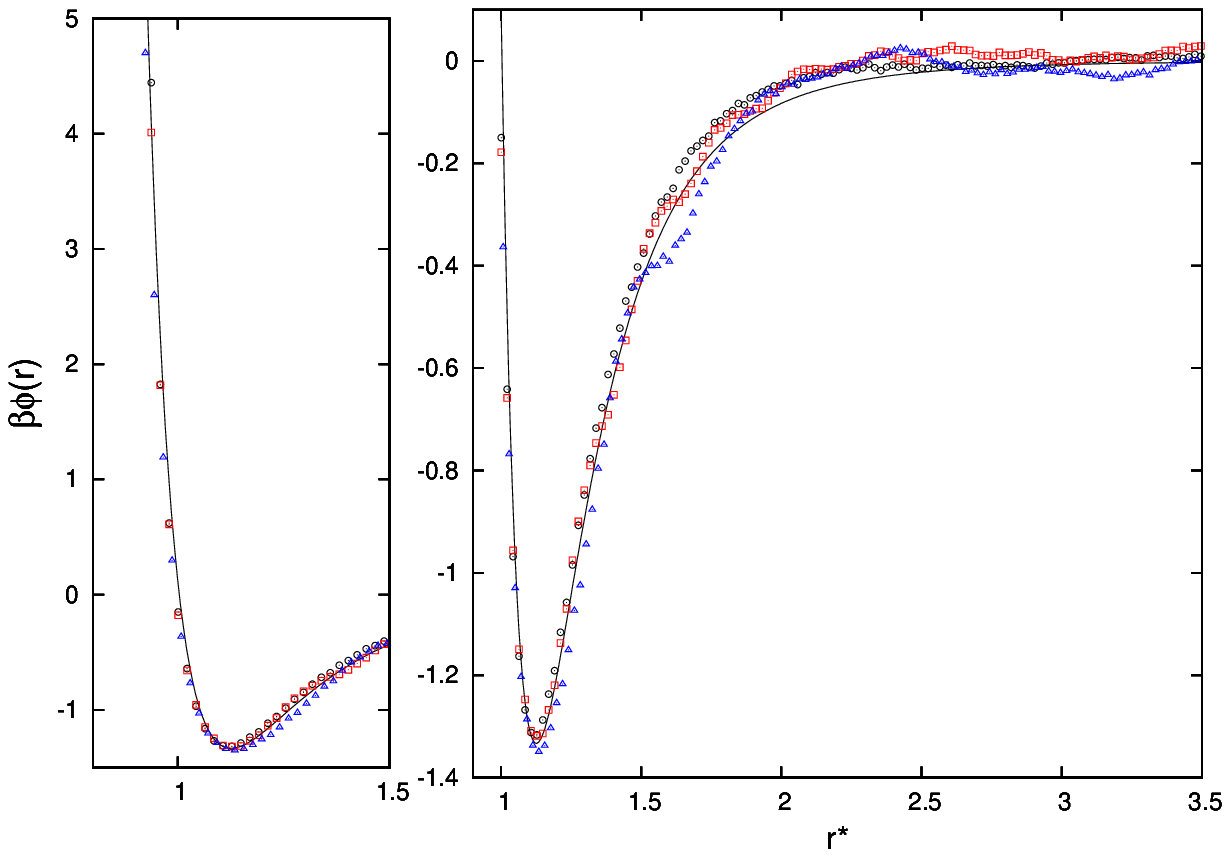}
\caption{\label{ResultLJpotential}(Color online) Results for the Lennard-Jones system. The target
potential (continuous line) and the model potential [circles for $g_{\sst{1}}(r)$, squares
for $g_{\sst{2}}(r)$ and triangles for $g_{\sst{3}}(r)$] are plotted.}
\end{figure}

In order to complete the inversion procedure we have evaluated the interaction potential associated to the three RDFs computed above.
These results have been obtained by using the method described in \cite{Dalessandro2010} and are presented in Fig. \ref{ResultLJpotential}.
The analysis of this figure shows that the potentials extracted from $g_{\sst{1}}(r)$ and $g_{\sst{2}}(r)$ are essentially equivalent and
provide a good estimate of the target one (with a maximum discrepancy of about $5\times 10^{-2}$ distributed over the $r$ axis).
The potential extracted from $g_{\sst{3}}(r)$ is less accurate with respect to the previous ones. In this case the main features of the target potential
(such as the amplitude and location of the absolute minimum) are reproduced correctly but some spurious oscillations are present. This behavior is a
consequence of the presence of small oscillations in the RDF $g_{\sst{3}}(r)$ which are hardly visible at the scale of Fig. \ref{SK-GR LJ}. This
fact indicates that a very precise reconstruction of the target RDF is needed in order to obtain a correct solution of the inverse problem.

\subsection{\label{ResultNa}Inversion of the Na data}

We present the result of the inversion of the structure factor of the liquid Na at 100 $^{\circ}$C \cite{Greenfield1971}.
Since we are dealing with a real fluid at high density we expect that the many-body contributions in the interaction potential
cannot be neglected, so our solution of the inverse problem will produce an effective pair potential.

Experimental data have been measured with a variable $k$ resolution up to 8.9 \AA ${}^{-1}$. In order to apply the inversion technique
defined in Sec. \ref{Theory} we need a target $S(k)$ sampled uniformly with a $\d k$ value compatible with the size of the simulation box;
so a preliminary operation on experimental data is needed.
Our prescription is the following: we perform the inverse Fourier transform of the experimental $S(k)$ and compute the (biased) RDF; then we
``cut'' this function at a value $r_{\sst{M}}$ consistent with the linear dimension of the model system in which we will perform the inversion procedure.
Finally, we transform back to the $k$ space and compute the new structure factor which is ready to be used as the target input function.
The reliability of this method has been tested for a Lennard-Jones system in which the evaluation of the target $S(k)$ and the inversion procedure for
the reconstruction of the RDF have been performed in boxes of different linear length. In all the cases we have obtained a correct reconstruction of
the target RDF. For the present case of the Na data we have chosen $r_{\sst{M}}=$ 22 \AA\ which corresponds to the maximum sampled value for a
model system made of 864 particles.

\begin{figure*}
  \begin{center}
    \begin{tabular}{cc}
      \resizebox{80mm}{!}{\includegraphics{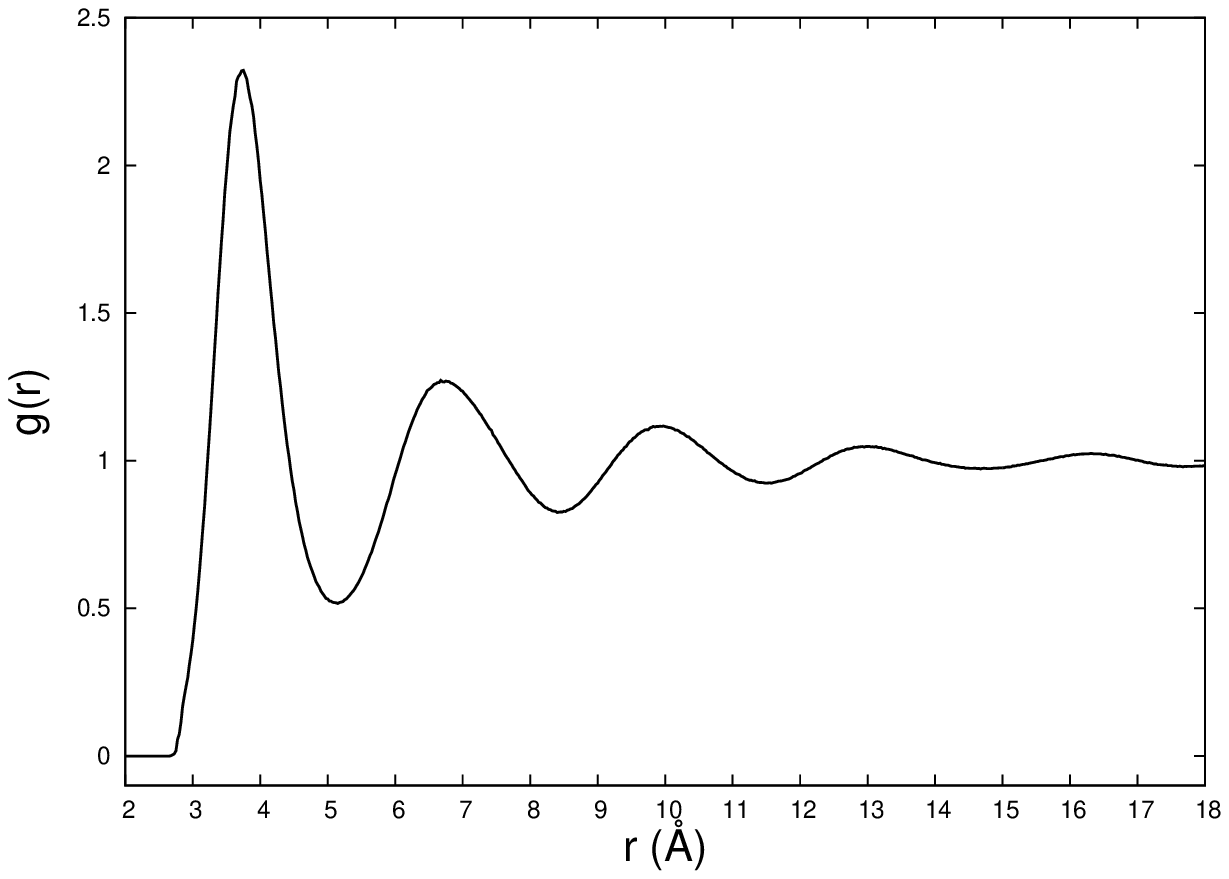}} &
      \resizebox{80mm}{!}{\includegraphics{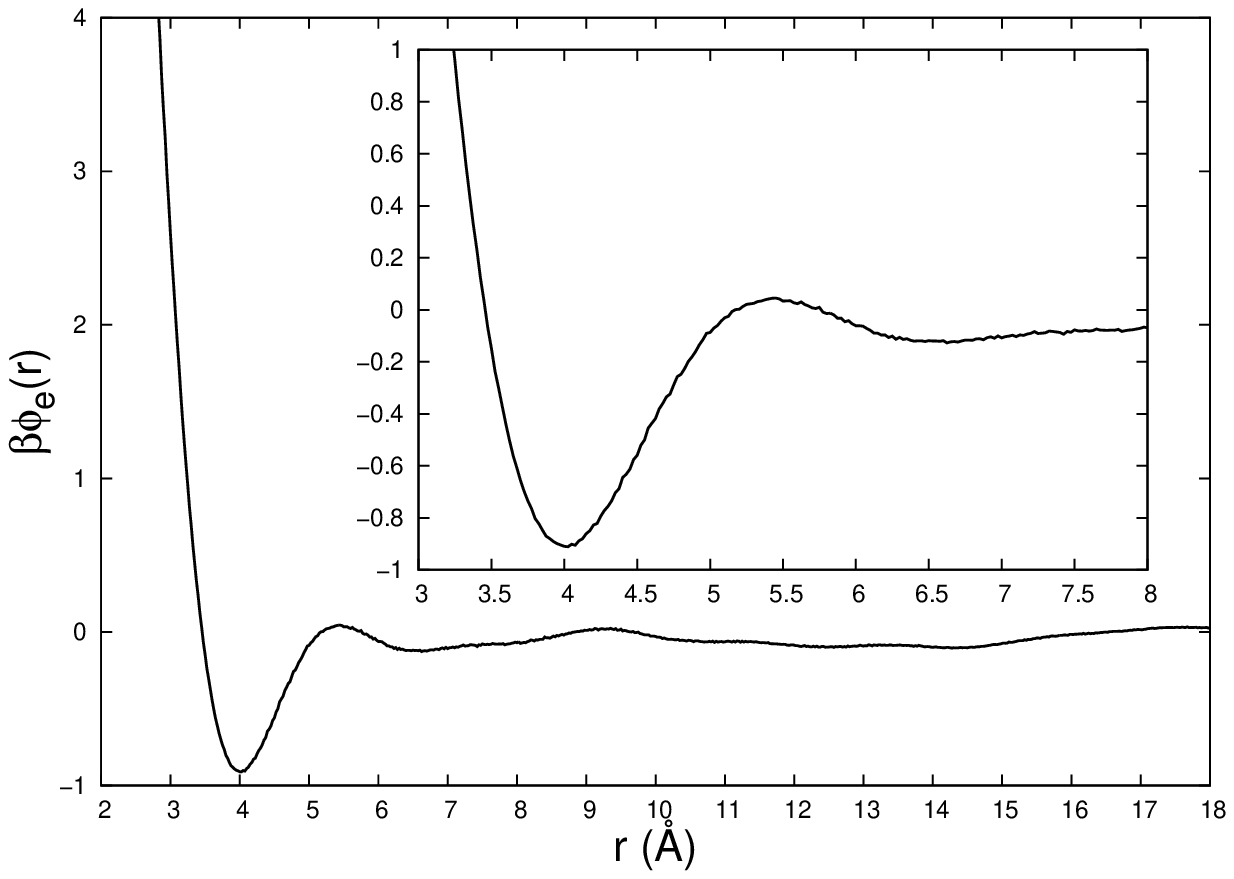}} \\
    \end{tabular}
    \caption{Results of the inversion procedure for Na  at 100$^{\circ}$C. The left panel contains the plot of the radial distribution function.
    The right panel contains the two-body effective potential.}
    \label{GR-PHI Na}
  \end{center}
\end{figure*}

The inverse simulation procedure for the reconstruction of the Na radial distribution function took $2\times 10^4$ cycles after equilibration.
The result for the RDF is reported in the left panel of Fig. \ref{GR-PHI Na}. This function evidences a HC radius of 2.65 \AA\ and the first peak
is located at $r=3.72$ \AA\ and is equal to 2.32. It is interesting to compare our result with the RDF obtained in \cite{Reatto1986} using an iterative
method for the inversion of the Na structure factor. The two functions are in substantial agreement: the RDF of \cite{Reatto1986} has a HC radius of
2.7 \AA, whereas the first peak is located at $r=3.66$ \AA\ and is equal to 2.43; furthermore even the relative positions of the other minima and maxima
differ less than the 2\%.

The Na pair effective potential has been extracted from the RDF computed above by using the method described in \cite{Dalessandro2010}. The result is
presented in the right panel of Fig. \ref{GR-PHI Na}. The potential reported in the figure evidences a highly repulsive part in the low $r$ region;
then there is an attractive zone with the minimum located in $r=$ 4.05 \AA\ and equal to $-0.91$ and a further weak repulsive part with a local maximum
at $r=$ 5.45 \AA . Finally, the potential approaches zero with some smooth oscillations. Again we compare our solution with the one obtained in
\cite{Reatto1986}. We observe that the shapes of the two potentials are in qualitative agreement but a quantitative comparison reveals some differences:
in particular, the locations of the absolute minima coincide but the depth of the potential wells differ by about 15\%. This fact has to be interpreted on
the basis of the high sensitivity of the inverse problem on the input RDF, so minor differences among the RDFs could produce a sensible discrepancy at the
level of the interaction potentials.

\section{\label{Discussion}Discussion and conclusions}

We have presented a method, based on the maximum entropy principle of information theory, for the reconstruction of the radial distribution
function of an equilibrium statistical system starting from the partial knowledge of its structure factor. The procedure is realized inside
a Monte Carlo simulation scheme which is revealed to be an effective tool for the implementation of the ME; indeed the maximization of the
configuration entropy is realized by the MC random displacements whereas the transition probability between neighbor states is defined
consistently with the information input codified in the target $S(k)$.
Once the RDF has been computed we can derive the two-body effective potential by using the method defined in \cite{Dalessandro2010},
thus providing a true ME-based solution of the inverse problem.

As stated in Sec. \ref{Max Ent approach}, the realization of the ME approach inside a MC-based procedure presents
some interesting features. Indeed, this method realizes a complete maximization of the model system configurational entropy (beyond
the two-body term) and provides the maximum entropy estimate of the complete equilibrium distribution of the model system.
So within this approach it is possible to extract informations concerning the physical system under inspection that goes beyond
the simple improvement of the Fourier transform of the structure factor.
Furthermore, since the correlators are obtained through the ensemble average over the model system configuration space, any non physical
feature (such as, for instance, negative values for the RDF inside the hard core region) is automatically avoided.

The applications of the method are presented in Sec. \ref{Applications}. Results analyzed in the first part of this section are designed to test
our approach with respect to the missing information in the input structure factor. ME has the feature of being ``maximally noncommittal
with regard to the missing information'' \cite{Jaynes1957}, and indeed, the results discussed in Sec. \ref{ResultLJ} demonstrate a reliable
reconstruction of the system RDF even for very limited knowledge of the $S(k)$.
Finally, Sec. \ref{ResultNa} contains the analysis of the real experimental data of the liquid sodium at 100 $^{\circ}$C. We evaluated the
Na RDF and then we extracted the effective pair interaction potential; both of the procedures converged to a stable result.
The solution of the inverse problem for this system has been compared with the one presented in \cite{Reatto1986}. The discrepancies between the
two potentials have been motivated in terms of the (small) differences among the RDFs. It is well known that the solution of the inverse problem
is highly sensible to the details of the pair function used as the input of the reconstruction procedure. So, under this perspective, the adoption
of the maximum entropy principle as a general and solid guideline for the definition inversion procedure could guarantee the correctness of the results.

A last comment concerns the possible extensions of the technique described in the present paper. ME principle holds for any system at equilibrium,
so the main idea at the basis of this approach can be extended to systems other than the simple monoatomic fluid discussed in the present paper.
For instance, polyatomic fluids are often characterized by strong directional interactions and an effective description of their physical properties
in terms of the model system defined in this paper could be revealed as very crude. In these cases, however, it is possible to define an improved model system
with new degrees of freedom which provide a better match with the ones of the experimental system under inspection. The statistical analysis
presented in Sec. \ref{Theory} has to be extended in order to include these new degrees of freedom and the same kind of procedure based on the
maximization of the log-likelihood ratio can be performed. Obviously, the feasibility of this strategy requires a higher involvement of information
concerning the target system and further experimental data, beyond the two-body pair function has to be provided.


\begin{acknowledgments}
The author is grateful to Francesco Cilloco for many stimulating discussions and suggestions.
A further acknowledgment goes to Luciana Silvestri for assistance.
\end{acknowledgments}


\bibliography{Reference}


\end{document}